\documentclass[twocolumn]{aastex631}

\usepackage{float} 
\usepackage{amsmath}
\usepackage{epsf}
\usepackage{color}
\usepackage{graphicx}
\usepackage{color}
\usepackage{enumitem}
\usepackage{mathtools}
\usepackage[mathcal]{eucal}
 \let\mathscr\relax
\usepackage[scr]{rsfso}

\shorttitle{First Look at  {\boldmath $z >$}~1 Bars in the Rest-Frame Near-IR  with {\it{JWST}} CEERS}
\shortauthors{Guo, Jogee, Finkelstein et al.}

\def\gtsim{\lower.5ex\hbox{$\buildrel > \over\sim$}}
\def\gtrsim{\mathrel{\hbox{\rlap{\hbox{\lower4pt\hbox{$\sim$}}}\hbox{$>$}}}}
\def\lesssim{\mathrel{\hbox{\rlap{\hbox{\lower4pt\hbox{$\sim$}}}\hbox{$<$}}}}
\def\ltsim{\lower.5ex\hbox{$\buildrel < \over\sim$}}
\def\simgt{{\raise-.5ex\hbox{$\buildrel>\over\sim$}}\ } 
\def\simlt{{\raise-.5ex\hbox{$\buildrel<\over\sim$}}\ }

\def\farcs{\hbox{$.\!\!^{\prime\prime}$}} 
 
\def\deg{{$^\circ$}}
\def\sun{\mbox{$_\odot$}}

\def\e#1{\underline{#1}}  

\begin{document}
\title{First Look at  {\boldmath $z >$}~1 Bars in the Rest-Frame
  Near-Infrared  with {\it{JWST}} Early CEERS Imaging}

\suppressAffiliations

\author[0000-0002-4162-6523]{Yuchen Guo}
\affiliation{Department of Astronomy, The University of Texas at Austin, Austin, TX, USA}

\author[0000-0002-1590-0568]{Shardha Jogee}
\affiliation{Department of Astronomy, The University of Texas at Austin, Austin, TX, USA}

\author[0000-0001-8519-1130]{Steven L. Finkelstein}
\affiliation{Department of Astronomy, The University of Texas at Austin, Austin, TX, USA}

\author{Zilei Chen}
\affiliation{Department of Astronomy, The University of Texas at Austin, Austin, TX, USA}

\author{Eden Wise}
\affiliation{Department of Astronomy, The University of Texas at Austin, Austin, TX, USA}

\author[0000-0002-9921-9218]{Micaela B. Bagley}
\affiliation{Department of Astronomy, The University of Texas at Austin, Austin, TX, USA}

\author[0000-0002-0786-7307]{Guillermo Barro}
\affiliation{Department of Physics, University of the Pacific, Stockton, CA 90340 USA}

\author[0000-0003-3735-1931]{Stijn Wuyts}
\affiliation{Department of Physics, University of Bath, Claverton
  Down, Bath BA2 7AY, UK}

\author[0000-0002-8360-3880]{Dale D. Kocevski}
\affiliation{Department of Physics and Astronomy, Colby College,
  Waterville, ME 04901, USA}

\author[0000-0001-9187-3605]{Jeyhan S. Kartaltepe}
\affiliation{Laboratory for Multiwavelength Astrophysics, School of Physics and Astronomy, Rochester Institute of Technology, 84 Lomb Memorial Drive, Rochester, NY 14623, USA}

\author[0000-0001-8688-2443]{Elizabeth J.\ McGrath}
\affiliation{Department of Physics and Astronomy, Colby College, Waterville, ME 04901, USA}

\author[0000-0001-7113-2738]{Henry C. Ferguson}
\affiliation{Space Telescope Science Institute, 3700 San Martin Dr., Baltimore, MD 21218, USA}

\author[0000-0001-5846-4404]{Bahram Mobasher}
\affiliation{Department of Physics and Astronomy, University of 
  California, 900 University Ave, Riverside, CA 92521, USA}

\author[0000-0002-7831-8751]{Mauro Giavalisco}
\affiliation{University of Massachusetts Amherst, 710 North Pleasant Street, Amherst, MA 01003-9305, USA}

\author[0000-0003-1581-7825]{Ray A. Lucas}
\affiliation{Space Telescope Science Institute, 3700 San Martin Dr., Baltimore, MD 21218, USA}

\author[0000-0002-7051-1100]{Jorge A. Zavala}
\affiliation{National Astronomical Observatory of Japan, 2-21-1 Osawa,
  Mitaka, Tokyo 181-8588, Japan} 

\author[0000-0003-3130-5643]{Jennifer M. Lotz}
\affiliation{Gemini Observatory/NSF's National Optical-Infrared Astronomy Research Laboratory, 950 N. Cherry Ave., Tucson, AZ 85719, USA}

\author[0000-0001-9440-8872]{Norman A. Grogin}
\affiliation{Space Telescope Science Institute, 3700 San Martin Dr., Baltimore, MD 21218, USA}

\author[0000-0002-1416-8483]{Marc Huertas-Company}
\affil{Instituto de Astrof\'isica de Canarias, La Laguna, Tenerife, Spain}
\affil{Universidad de la Laguna, La Laguna, Tenerife, Spain}
\affil{Universit\'e Paris-Cit\'e, LERMA - Observatoire de Paris, PSL, Paris, France}

\author[0000-0003-2338-5567]{Jes\'us Vega-Ferrero}
\affil{Instituto de Astrof\'isica de Canarias, La Laguna, Tenerife, Spain}

\author[0000-0001-6145-5090]{Nimish P. Hathi}
\affiliation{Space Telescope Science Institute, 3700 San Martin Dr., Baltimore, MD 21218, USA}

\author[0000-0002-7959-8783]{Pablo Arrabal Haro}
\affiliation{NSF's National Optical-Infrared Astronomy Research Laboratory, 950 N. Cherry Ave., Tucson, AZ 85719, USA}

\author[0000-0001-5414-5131]{Mark Dickinson}
\affiliation{NSF's National Optical-Infrared Astronomy Research Laboratory, 950 N. Cherry Ave., Tucson, AZ 85719, USA}

\author[0000-0002-6610-2048]{Anton M. Koekemoer}
\affiliation{Space Telescope Science Institute, 3700 San Martin Dr., Baltimore, MD 21218, USA}

\author[0000-0001-7503-8482]{Casey Papovich}
\affiliation{Department of Physics and Astronomy, Texas A\&M University, College Station, TX, 77843-4242 USA}
\affiliation{George P.\ and Cynthia Woods Mitchell Institute for Fundamental Physics and Astronomy, Texas A\&M University, College Station, TX, 77843-4242 USA}

\author[0000-0003-3382-5941]{Nor Pirzkal}
\affiliation{ESA/AURA Space Telescope Science Institute}

\author[0000-0003-3466-035X]{L. Y. Aaron\ Yung}
\altaffiliation{NASA Postdoctoral Fellow}
\affiliation{Astrophysics Science Division, NASA Goddard Space Flight Center, 8800 Greenbelt Rd, Greenbelt, MD 20771, USA}

\author[0000-0001-8534-7502]{Bren E. Backhaus } 
\affil{Department of Physics, 196A Auditorium Road, Unit 3046, University of Connecticut, Storrs, CT 06269, USA}

\author[0000-0002-5564-9873]{Eric F.\ Bell}
\affiliation{Department of Astronomy, University of Michigan, 1085 S. University Ave, Ann Arbor, MI 48109-1107, USA}

\author[0000-0003-2536-1614]{Antonello Calabr\`o}
\affiliation{INAF Osservatorio Astronomico di Roma, Via Frascati 33, 00078 Monteporzio Catone, Rome, Italy}

\author[0000-0001-7151-009X]{Nikko J. Cleri}
\affiliation{Department of Physics and Astronomy, Texas A\&M University, College Station, TX, 77843-4242 USA}
\affiliation{George P.\ and Cynthia Woods Mitchell Institute for Fundamental Physics and Astronomy, Texas A\&M University, College Station, TX, 77843-4242 USA}

\author[0000-0002-4343-0479]{Rosemary T. Coogan}
\affiliation{CEA, IRFU, DAp, AIM, Universit\'{e} Paris-Saclay, Universit\'{e} Paris Cit\'{e}, Sorbonne Paris Cit\'{e}, CNRS, 91191 Gif-sur-Yvette, France}

\author[0000-0003-1371-6019]{M. C. Cooper}
\affiliation{Department of Physics \& Astronomy, University of California, Irvine, 4129 Reines Hall, Irvine, CA 92697, USA}

\author[0000-0001-6820-0015]{Luca Costantin}
\affiliation{Centro de Astrobiolog\'ia (CSIC-INTA), Ctra de Ajalvir km 4, Torrej\'on de Ardoz, 28850, Madrid, Spain}

\author[0000-0002-5009-512X]{Darren Croton}
\affiliation{Centre for Astrophysics \& Supercomputing, Swinburne University of Technology, Hawthorn, VIC 3122, Australia}
\affiliation{ARC Centre of Excellence for All Sky Astrophysics in 3 Dimensions (ASTRO 3D)}

\author[0000-0001-8047-8351]{Kelcey Davis}
\affiliation{Department of Physics, 196 Auditorium Road, Unit 3046, University of Connecticut, Storrs, CT 06269, USA}

\author[0000-0002-6219-5558]{Alexander de la Vega}
\affiliation{Department of Physics and Astronomy, University of California, Riverside, CA 92521, USA}

\author[0000-0003-4174-0374]{Avishai Dekel}
\affil{Racah Institute of Physics, The Hebrew University of Jerusalem,  Jerusalem 91904, Israel}

\author[0000-0002-3560-8599]{Maximilien Franco}
\affiliation{Department of Astronomy, The University of Texas at Austin, Austin, TX, USA}

\author[0000-0003-2098-9568]{Jonathan P. Gardner}
\affiliation{Astrophysics Science Division, NASA Goddard Space Flight Center, 8800 Greenbelt Rd, Greenbelt, MD 20771, USA}

\author[0000-0002-4884-6756]{Benne W. Holwerda}
\affil{Physics \& Astronomy Department, University of Louisville, 40292 KY, Louisville, USA}

\author[0000-0001-6251-4988]{Taylor A. Hutchison}
\altaffiliation{NASA Postdoctoral Fellow}
\affiliation{Astrophysics Science Division, NASA Goddard Space Flight Center, 8800 Greenbelt Rd, Greenbelt, MD 20771, USA}

\author[0000-0002-2499-9205]{Viraj Pandya}
\altaffiliation{Hubble Fellow}
\affiliation{Columbia Astrophysics Laboratory, Columbia University, 550 West 120th Street, New York, NY 10027, USA}

\author[0000-0003-4528-5639]{Pablo G. P\'erez-Gonz\'alez}
\affiliation{Centro de Astrobiolog\'{\i}a (CAB), CSIC-INTA, Ctra. de Ajalvir km 4, Torrej\'on de Ardoz, E-28850, Madrid, Spain}

\author[0000-0002-5269-6527]{Swara Ravindranath}
\affiliation{Space Telescope Science Institute, 3700 San Martin Dr., Baltimore, MD 21218, USA}

\author[0000-0002-8018-3219]{Caitlin Rose}
\affil{Laboratory for Multiwavelength Astrophysics, School of Physics and Astronomy, Rochester Institute of Technology, 84 Lomb Memorial Drive, Rochester, NY 14623, USA}

\author[0000-0002-1410-0470]{Jonathan R. Trump}
\affil{Department of Physics, 196A Auditorium Road, Unit 3046, University of Connecticut, Storrs, CT 06269, USA}

\author[0000-0002-9593-8274]{Weichen Wang}
\affiliation{Department of Physics and Astronomy, Johns Hopkins
  University, 3400 N. Charles Street, Baltimore, MD 21218, USA}

\begin{abstract}
Stellar bars are key drivers of secular evolution in galaxies and
can be effectively studied using rest-frame near-infrared (NIR) images, which
trace the underlying stellar mass and are less impacted by dust and
star formation than rest-frame UV or optical images. 
We leverage the power of {\it{JWST}} CEERS NIRCam images 
to present the first quantitative identification and characterization of
stellar bars  at $z>1$ based on rest-frame NIR F444W images of high
resolution ($\sim$1.3 kpc  at $z\sim$~1--3).
We identify stellar bars in these images using quantitative criteria based on ellipse
fits. For this pilot study, we present six examples of robustly identified 
bars at  $z>1$ with spectroscopic redshifts, including the two highest redshift bars  at  $z\sim$~2.136  and
2.312 quantitatively identified and characterized to date.
The stellar bars at $z\sim$~1.1--2.3  presented in our study
have projected  
semi-major axes of  $\sim$~2.9--4.3 kpc  and 
projected ellipticities of $\sim$~0.41--0.53 
in the rest-frame NIR.  
The barred host galaxies have stellar masses  $\sim 1 \times
10^{10}$  to $2 \times 10^{11}$ $M_{\odot}$, 
star formation rates of  $\sim$ 21--295 $M_{\odot}$ yr$^{-1}$, and 
several have potential nearby companions. 
Our finding of bars at $z\sim$~1.1--2.3 demonstrates  the early onset
of such instabilities and supports simulations  
where  bars form early  in massive dynamically cold 
disks. It also suggests that if these bars at lookback times of 8--10
Gyr  survive out to present epochs, bar-driven secular processes may operate  over a long time 
and have a significant impact on some galaxies  by $z \sim$~0.
\end{abstract} 


\section{Introduction}\label{sec:intro}

Stellar bars play a central role in the secular evolution of galaxies
by  efficiently redistributing mass and angular momentum and 
driving gas inflows into the circumnuclear region through gravitational torques
and shocks (e.g., \citealt{Athanassoula2002, Athanassoula-Lambert-Dehnen2005, Kormendy-Kennicutt2004, Jogee-Scoville-Kenney2005}). Most present-day spirals are barred (e.g., \citealt{Eskridge-etal-2000, Laurikainen-Salo-Buta2004, Marinova-Jogee2007, Menendez-Delmestre-etal-2007}), including our own Milky Way (\citealt{Peters1975, Blitz-Spergel1991, Binney-etal-1991,Weiland-etal-1994}).
 
Observational evidence in nearby galaxies suggests bars influence their central molecular gas concentrations (e.g.,
\citealt{Sakamoto-etal-1999, Jogee-Scoville-Kenney2005}), velocity fields of ionized gas
(e.g., \citealt{Regan-Vogel-Teuben1997}),  star formation (SF) activity (e.g.,
\citealt{Hunt-Malkan1999, Jogee-Scoville-Kenney2005, Masters-etal-2010, George-Subramanian2021}), 
and central bulges (e.g., \citealt{Kormendy-Kennicutt2004,Jogee-Scoville-Kenney2005,Gadotti-etal-2015}). 
The role of bars on AGN is less clear as both
simulations (e.g., \citealt{Combes-Gerin1985, Athanassoula1992b}) and observations
(e.g., \citealt{Knapen-etal-1995, Buta-Combes1996, Jogee-Scoville-Kenney2005}) show
that bar-driven gas inflows tend to stall in the circumnuclear
region where the specific angular momentum of the gas is still too high to fuel the AGN (\citealt{Jogee2006}).

The exploration of stellar bars out to early cosmic times is important for understanding the growth and morphological transformation of galaxies, a process which is driven since $z\sim$~4 by
gas accretion \citep[e.g.,][]{Katz-etal-2003, Keres-etal-2005, Keres-etal-2012, Dekel-Birnboim2006,Faucher-Giguere-Keres2011}, galaxy mergers and tidal interactions
\citep[e.g.,][]{Conselice-etal-2003,Kartaltepe-etal-2007, Jogee-etal-2009,Lotz-etal-2010},
as well as bar-driven secular processes.
Numerous studies show near-infrared (NIR) images are better tracers
than optical images of the stellar mass distribution and structural
components of galaxies as the effects of dust extinction and  SF 
are lower in the NIR (e.g., \citealt*{Frogel-Quillen-Pogge1996}, \citealt{Suess-etal-2022}),
and the mass-to-light ratio in the NIR is less sensitive to the ages
of the stellar populations (e.g., \citealt{Schneider2006}).  Indeed, 
the bar fraction in bright spirals at $z\sim$~0 is higher
in the NIR (e.g., \citealt{Marinova-Jogee2007,Menendez-Delmestre-etal-2007}) than in the optical.
However, to date, studies of bars out to  $z\sim$~1 have only been able to use
the rest-frame optical light traced by {\it{Hubble Space Telescope (HST)}}
WFPC2, ACS, NICMOS, and WFC3 images.

Early  {\it{HST}} studies of bars in the rest-frame optical by
\cite{Elmegreen-Elmegreen-Hirst2004} and \cite{Jogee-etal-2004} presented the 
first evidence of a significant population of barred galaxies  out to
$z \sim$~1 (lookback time of $\sim$~8 Gyr), showing 
that bars are already in place at early times and implying that 
bar-driven secular processes can potentially operate 
over many billions of years if these bars survive to the present day.
Results on how the bar fraction varies out to $z \sim$~1 have been
mixed: some studies (e.g., \citealt{Elmegreen-Elmegreen-Hirst2004,Jogee-etal-2004}) do not
find a strong decline in the bar fraction out to  $z \sim$~1,
other studies find a decline by a factor of a few
(e.g., \citealt{Sheth-etal-2008, Melvin-etal-2014}), while \cite{ Cameron-etal-2010}
points out that results on the bar fraction depend on the stellar mass range.

The vast majority of  {\it{HST}} studies have explored  bars in the rest-frame optical light 
out to $z\sim$~1.2 (e.g., \citealt{Abraham-etal-1999, Elmegreen-Elmegreen-Hirst2004,
  Jogee-etal-2004, Sheth-etal-2008, Cameron-etal-2010, Melvin-etal-2014}).  The 
study by \cite{Simmons-etal-2014} represented a first attempt to push the
explorations of bars in the rest-frame optical  out to $z \le$~2, but faced challenges in robustly
characterizing bars at $z >$~1.5.

The advent of sensitive, high-resolution NIRCam images from  the
{\it{James Webb Space Telescope}} ({\it{JWST}}; \citealt{Gardner-etal-2006})
holds the promise of tremendous advances in the exploration of bars at
$z >$~1 and provides us  for the first time with 
high-resolution rest-frame NIR images at $z >$~1. At the same time, 
new high-resolution  cosmological simulations
(e.g., \citealt*{Kraljic-Bournaud-Martig2012}; \citealt{Scannapieco-Athanassoula2012,Bonoli-etal-2016,Spinoso-etal-2017,Algorry-etal-2017,Fragkoudi-etal-2021,Rosas-Guevara-etal-2020,
  Rosas-Guevara-etal-2022,Bi-Shlosman-Romano-Diaz2022}) are probing the growth of bars and their impact
on galaxy evolution out to $z \ge$~4.

In this pilot study we conduct the first quantitative exploration of  stellar bars
at $z>1$  in high-resolution rest-frame NIR images 
by analyzing {\it{JWST}} NIRCam images 
in the first epoch of imaging from the
Cosmic Evolution Early Release Science Survey (CEERS; \citealt{Finkelstein-etal-2022}). 
Thanks to the {\it{JWST}} F444W images, we can
for the first time use high-resolution (0\farcs16 corresponding to $\sim$1.3 kpc  at
$z\sim$~1--3) rest-frame NIR images to quantitatively identify and characterize bars at  $z>1$.
The sample selection is outlined in \S~\ref{sec:sample}.
In \S~\ref{sec:method} we describe our methodology to identify 
and characterize bars based on the application of  physically motivated quantitative criteria to ellipse fits.
For this pilot study, we present in \S~\ref{sec:resul} six examples of robustly identified 
bars at  $z>1$ with spectroscopic redshifts, including the two highest redshift bars  at  $z\sim$~2.136  and
2.312 quantitatively identified and characterized to date.
\S~\ref{sec:discu} discusses the implications of our results for the onset and impact of early generations of
bars on galaxy evolution.

We stress that this pilot study only presents six examples  of robustly
identified bars at  $z>1$ in the rest-frame NIR  rather than 
a full census of all observable bars at $z >$~1. 
In  future papers that will incorporate the upcoming 
additional six CEERS pointings, we will present such a full 
census of observable bars at $z>1$, estimate the rest-frame optical and 
NIR bar fraction, and explore the relationship between 
bars and galaxy properties (SF, bulges, AGN, and presence of
companions) using a control sample of unbarred galaxies.

In this paper we assume the latest {\it Planck} flat $\Lambda$CDM cosmology with H$_{0}=$67.36, $\Omega_m=$0.3153, and $\Omega_{\Lambda}=$0.6847 \citep{Planck-etal-20}.  All magnitudes are in the absolute bolometric system \citep[AB; ][]{Oke-Gunn1983}.

\section{CEERS Observations and Data Reduction}\label{sec:ceers}

CEERS is one of 13 early release science surveys designed to obtain data covering all areas of astronomy early in Cycle 1. In this
pilot paper we use the  first epoch of CEERS NIRCam imaging, which has four of the planned ten pointings obtained on 21 June 2022, known
as CEERS1, CEERS2, CEERS3, and CEERS6. We refer the reader to the
CEERS survey \citep{Finkelstein-etal-2022} and  data reduction (\citealt{Bagley-etal-2022}) papers for a full description of the CEERS survey and briefly summarize the key aspects here.
Data were obtained  in each pointing in  the short-wavelength (SW)
channel F115W, F150W, and F200W filters, and long-wavelength (LW)
channel F277W, F356W, F410M, and F444W filters with a typical exposure
time of  2835 s per filter in each of three dithers, except for F115W
which had longer exposure times.  A careful initial reduction of the NIRCam images in all
four pointings was performed using version 1.5.3 of the \textit{JWST} Calibration Pipeline\footnote{\url{jwst-pipeline.readthedocs.io}} with some
custom modifications. Version v0.07 of the CEERS data reduction was used in this work.
As described in \cite{Finkelstein-etal-2022}, data were processed through Stages 1 and 2 of the pipeline 
where reduction steps included detector-level correction, wisp subtraction,  removal of \textit{1/f} noise, 
flat fielding, and masking of bad pixels.
This was followed by astrometric correction and co-addition of calibrated detector images 
 onto a common output grid using the drizzle algorithm 
with an inverse variance map weighting \citep{Casertano-etal-2000,Fruchter-Hook02}.
The RMS of the absolute alignment to {\it{HST}} F160W is $\sim$~12-15 mas and the RMS of the NIRCam-to-NIRCam alignment is $\sim$~5-10 mas.
The output mosaics have pixel scales of 0\farcs03/pixel. 
The usable total area covered by these observations is 34.5 arcmin$^2$.
As described in \cite{Finkelstein-etal-2022} the CEERS v0.07 photometry catalog was produced by using \textsc{Source Extractor} \citep[][]{Bertin-Arnouts96}  v2.25.0 in two image mode, with an inverse-variance weighted combination of the PSF-matched F277W and F356W images as the detection image, and photometry measured on all seven bands.

\section{Sample Selection}\label{sec:sample}

For this pilot study we follow the procedure below to  identify a sample of galaxies with stellar mass $M_{\ast} \ge 10^{10} M_{\odot}$ at  redshifts 
1~$\le z \le $~3 that are  in the Multi-wavelength Catalogs
for the Extended Groth strip \citep[EGS;][]{Stefanon-etal-2017} for CANDELS \citep{Grogin-etal-2011,Koekemoer-etal-2011} and have
CEERS NIRCam imaging. 

The redshift range 1~$\le z \le $~3  was selected for the following reasons.
We set an upper limit of $z \le $~3 so that we can trace  the
rest-frame NIR light at wavelengths  $\lambda \ge 1.1$~microns  using  the
longest-wavelength F444W NIRCam image.  
We set our lower limit at $z \ge$~1 as most {\it{HST}} studies have explored  bars in the
rest-frame optical light  out to $z\sim$~1.2 (see  \S~\ref{sec:intro})
and the properties of bars at  $z > $~1  constitute an
uncharted territory of great interest. Additionally, in the redshift
range of $z \sim$~1--3, the empirically measured point spread function
(PSF) of  0\farcs16 in the F444W band 
corresponds to a high spatial resolution of $\sim$~1.3 kpc.

The  sample of galaxies  is derived by cross-matching the CEERS v0.07 source
catalog  \citep{Finkelstein-etal-2022} with 
 the CANDELS EGS catalog
 \citep{Stefanon-etal-2017} within  0\farcs25, and identifying 
galaxies with stellar mass $M_{\ast} \ge 10^{10} M_{\odot}$ at  redshifts 
1~$\le z \le $~3.  
At $z \sim$ 2, the 90\% stellar mass completeness of the CANDELS EGS catalog is $\sim$ $10^{10}
M_{\odot}$ \citep{Stefanon-etal-2017}.   
This cross-matching results in a sample of 
348 galaxies
with  stellar mass $M_{\ast} \ge 10^{10} M_{\odot}$ at  redshifts 
1~$\le z \le $~3. 

We use the robust photometric redshifts and stellar mass measurements
in the value-added catalogs associated with the CANDELS EGS catalogs
\citep{Stefanon-etal-2017}.  For $\sim$~67\% of the  sample, we supplement
photometric redshifts  with available published spectroscopic
redshifts in EGS (N. Hathi 2022, private communication). 
If a source has more than one spectroscopic redshift measurement, we choose the
one with the highest quality.

\section{Methodology}\label{sec:method}
\subsection{Identification of Stellar Bars}\label{sec:method-identification}

\begin{figure*}[!t]
\centering
 \includegraphics[width=\textwidth]{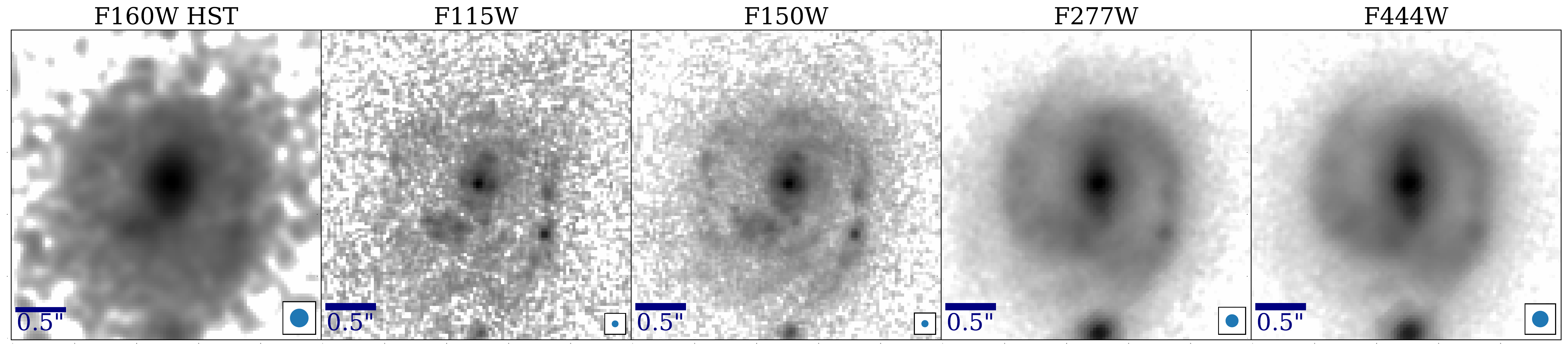} 
\vspace{0mm}
\caption{
This figure illustrates the effects of bandpass shift and PSF for  an example barred
galaxy (EGS-23205) at redshift $z\sim$~2.136 in our sample.
From left to right, we show the 
{\it{HST}}  WFC3 F160W, and {\it{JWST}} NIRCam F115W, F150W,
F277W, and F444W images.
The blue circle at the bottom right of each image represents the point
spread function (PSF) FWHM of each band (0\farcs18, 0\farcs07,
0\farcs07, 0\farcs13, and 0\farcs16, respectively)  and the horizontal bar shows a 0\farcs5 scale for reference. All images are 3\farcs0 $\times$ 3\farcs0 in size.
The underlying stellar mass distribution and galactic components, such
as the stellar bar, are better traced by the high-resolution
rest-frame NIR image revealed by the {\it{JWST}} F444W data than by
the rest-frame UV light shown in the {\it{JWST}} F115W images. 
It is also striking that although the  {\it{HST}} F160W and  {\it{JWST}} F444W images
have a similar PSF (0\farcs18 and  0\farcs16, respectively),
the bar is more evident in the JWST image due to the longer
rest-frame wavelength light the latter is tracing.
Signs of the bar are also visible in the high-resolution rest-frame
red optical image  traced by the {\it{JWST}} F227W data,
but are much less evident  in the rest-frame blue optical light traced by 
the {\it{JWST}} F150W image.
In all images, $N$ is up and $E$ is left.
}
\label{fig:bands}
\end{figure*}

Stellar bars are  non-axisymmetric, flattened triaxial systems within 
stellar disks that are made up of families of periodic stellar orbits
that conserve the Jacobi  integral.
The  main bar-supporting  family of  {\bf{x}}$_1$ orbits  are elongated
along the long-axis of the stellar  bar (e.g., \citealt{
Contopoulos-Papayannopoulos1980, Athanassoula1992a}).
Our methodology to identify bars in the {\it{JWST}} data consists of
two stages outlined below.


{\bf{Stage 1:}}
The first stage is a liberal visual classification whose goal is to cast as wide a
net as possible for systems with elongated structures that may be
putative bar candidates, with the idea that subsequent ellipse fits of 
these candidates would allow us to identify the barred systems.
For the visual classification,  we visually inspected
postage stamps of six NIRCam images
(F115W, F150W,  F200W, F277W, F356W, and F444W)
of our 348 sample galaxies at  1~$\le z \le $~3
to first remove unresolved systems and 
very strongly distorted and asymmetric systems. 
Among the remaining galaxies, we then liberally selected 
a sample S1 of galaxies that  host any elongated structures (in any
band)  that might even marginally be stellar bars. 
We ended up with 82 galaxies in sample S1.

{\bf{Stage 2:}}
The second stage involves using the methodology described in
\cite{Jogee-etal-2004} and \cite{Marinova-Jogee2007}  
to identify bars. In brief, this methodology involves ellipse-fitting the tracer images 
(e.g., \citealt{Jedrzejewski1987, Wozniak-etal-1995, Jogee-etal-2002a, Jogee-etal-2004,Elmegreen-Elmegreen-Hirst2004,Marinova-Jogee2007}), followed by the application of quantitative  criteria to identify bars. 
For this second stage, we ellipse fitted the F444W image of each
galaxy in sample S1. 
Figure~\ref{fig:bands}  illustrates the effects of bandpass shift and PSF for  an example barred
galaxy (EGS-23205) at redshift $z\sim$~2.136 in our sample.
The stellar bar is evident in the high-resolution rest-frame NIR 
({\it{JWST}} F444W) image, but not in the rest-frame UV ({\it{JWST}} F115W) image. 
The bar is more evident in the rest-frame NIR {\it{JWST}} F444W image 
than in the rest-frame blue optical  {\it{HST}} F160W  image although
the images have a similar PSF (0\farcs18 and  0\farcs16, respectively).

\begin{figure*}[h]
\centering

\includegraphics [width=0.67 \textwidth]{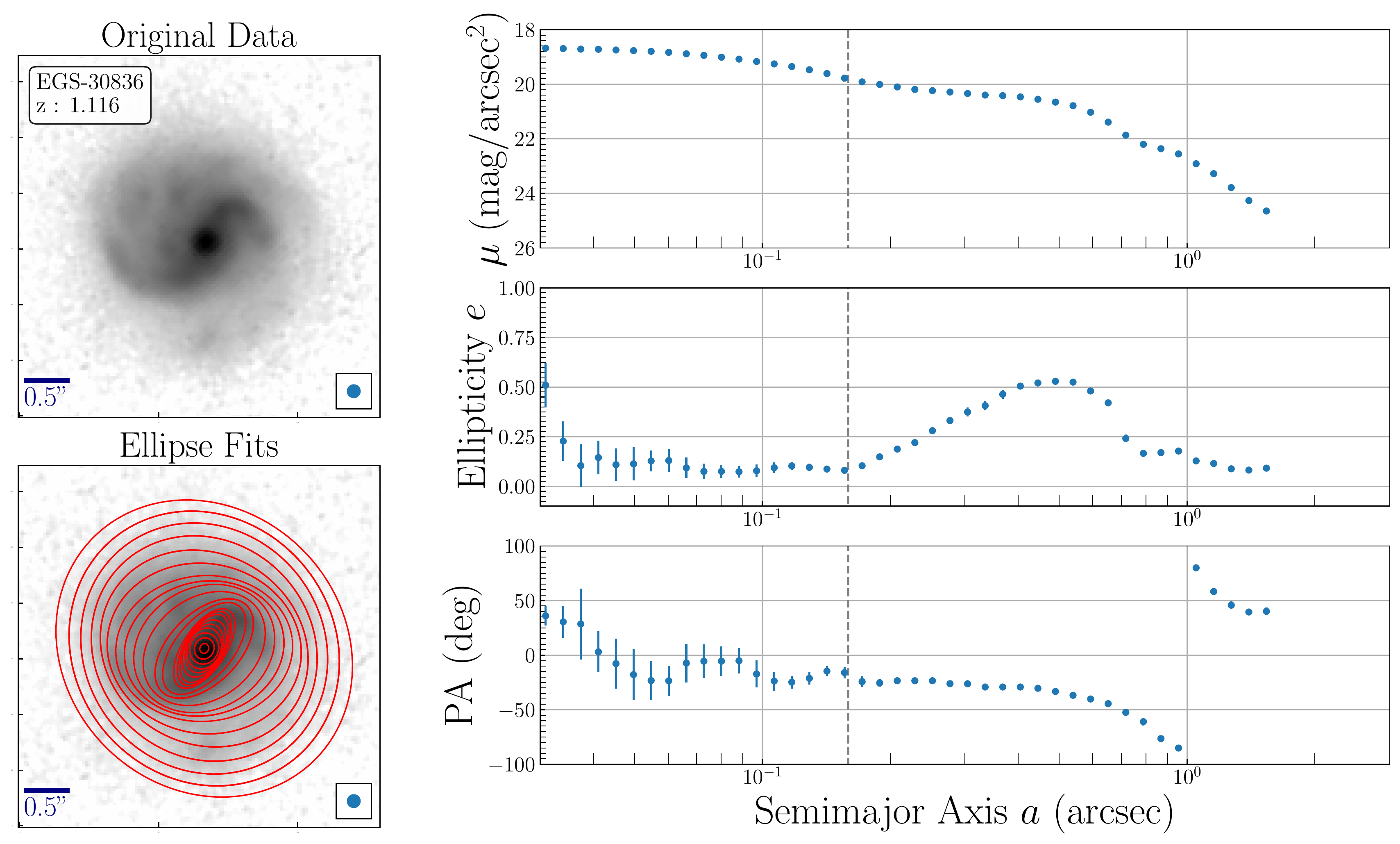}
\includegraphics [width=0.67\textwidth]{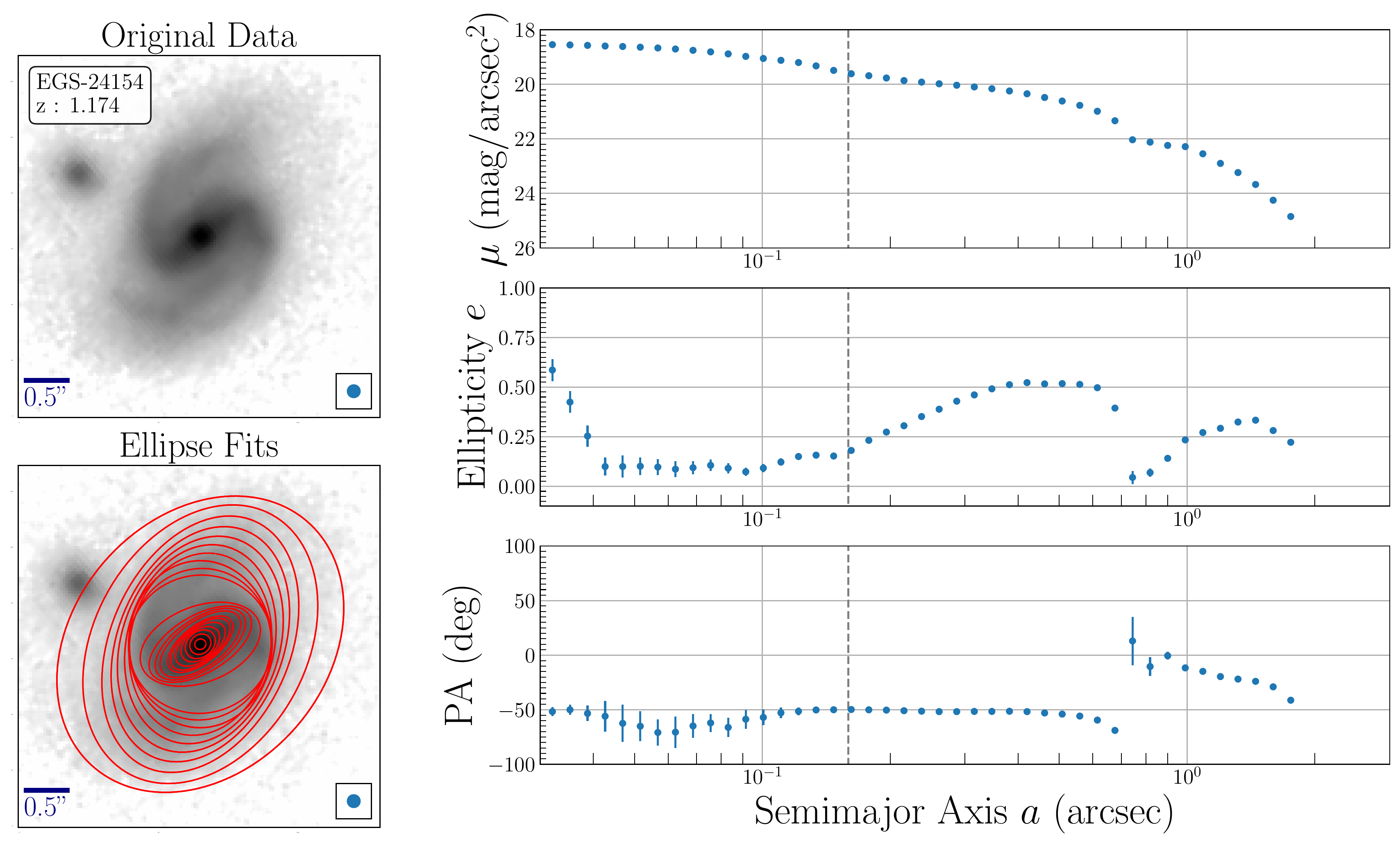}
\includegraphics [width=0.67 \textwidth]
{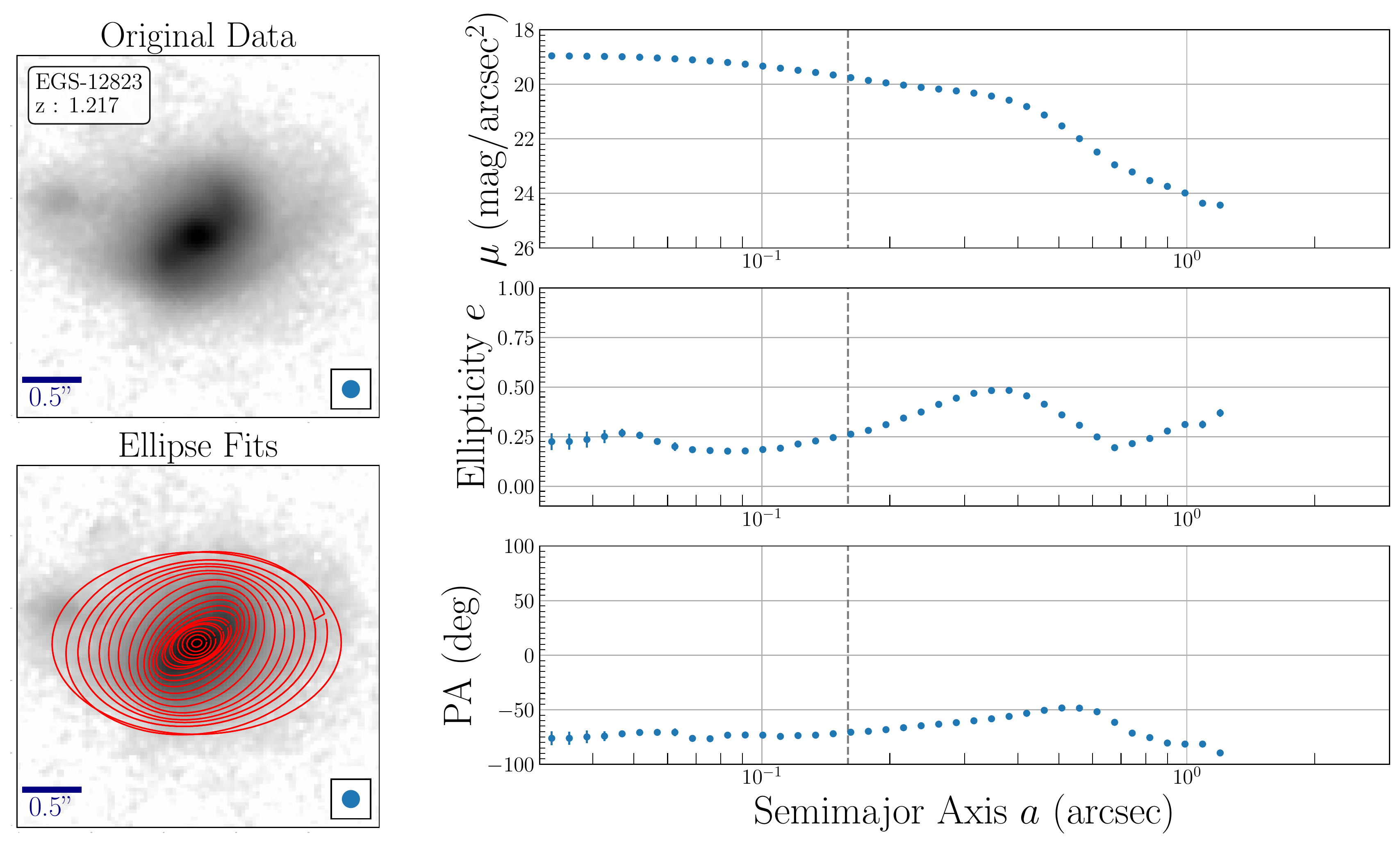}
\vspace{-1mm}
\caption{
Ellipse fits to the {\it{JWST}} NIRCam F444W image of 
three example  barred galaxies (EGS-30836, EGS-24154, EGS-12823). 
The left panel for each galaxy shows 
the F444W image alone (top) and then with the ellipse fits superposed
(bottom). Although nearby sources may appear in the images, they 
are masked during the ellipse fitting.
The blue circle at the bottom right of each image represents
the PSF FWHM (0\farcs16 corresponding to $\sim$1.3 kpc  at
$z\sim$~1--3), and the horizontal bar shows  a 0\farcs5 scale for reference. Size of each image is adjusted with respect to the size of the source, and ranges from 3\farcs0 $\times$ 3\farcs0 to 3\farcs9 $\times$ 3\farcs9.
The right panel for each galaxy shows
the radial profiles of surface brightness ($\mu$), ellipticity ($e$),
and position angle (PA)  versus semi-major axis $a$ derived from the
ellipse fits. See \S~\ref{sec:method-identification} for details.
PA goes from 0 to -90 clockwise (from North to West) and goes from 0
to 90 counter-clockwise (from North to East). The vertical dashed line
represents the F444W PSF (0\farcs16). 
}
\label{fig:ellip1}
\end{figure*}

\begin{figure*}[h]
\centering
\includegraphics[width=0.7\textwidth]{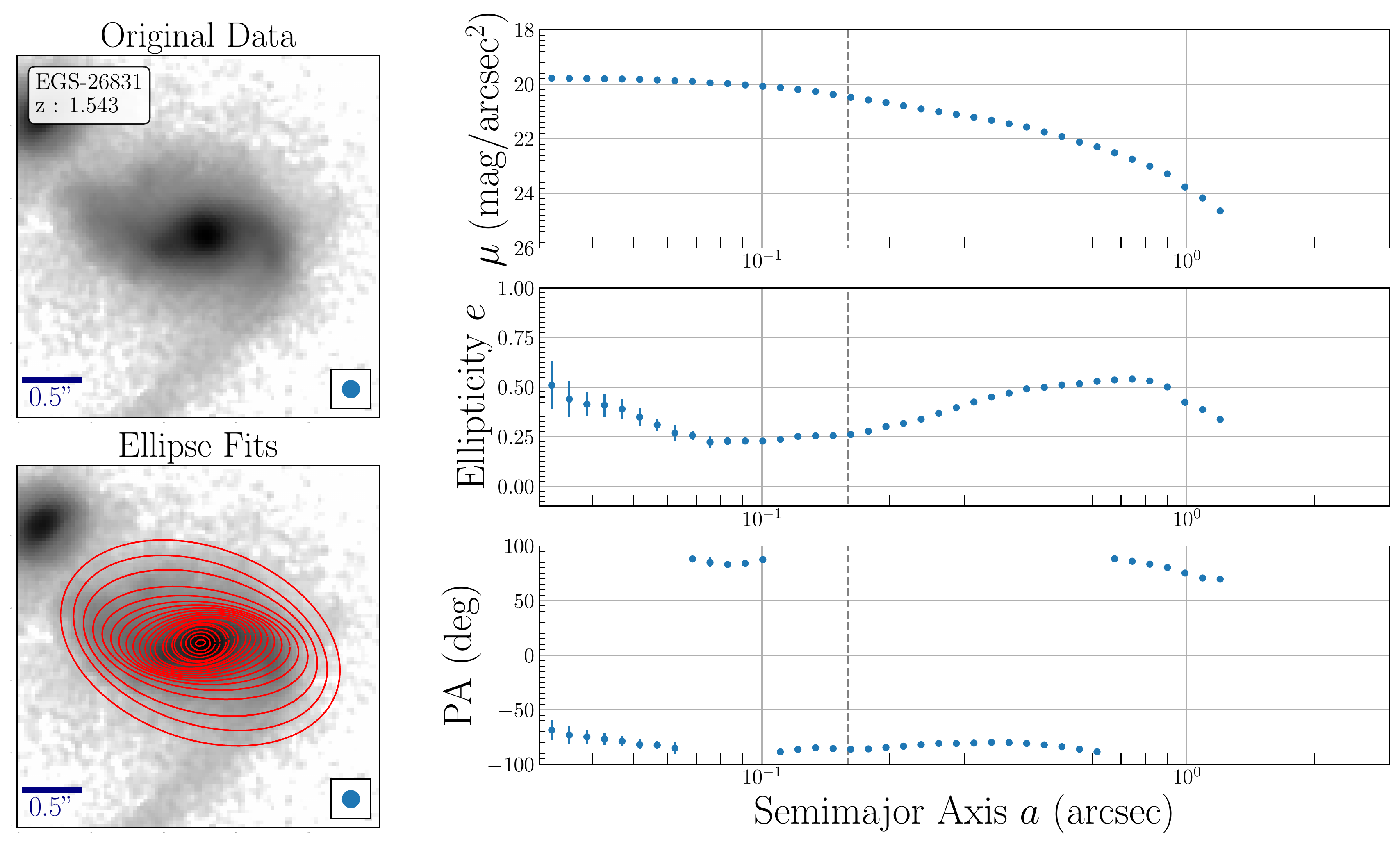}
\includegraphics[width=0.7\textwidth]{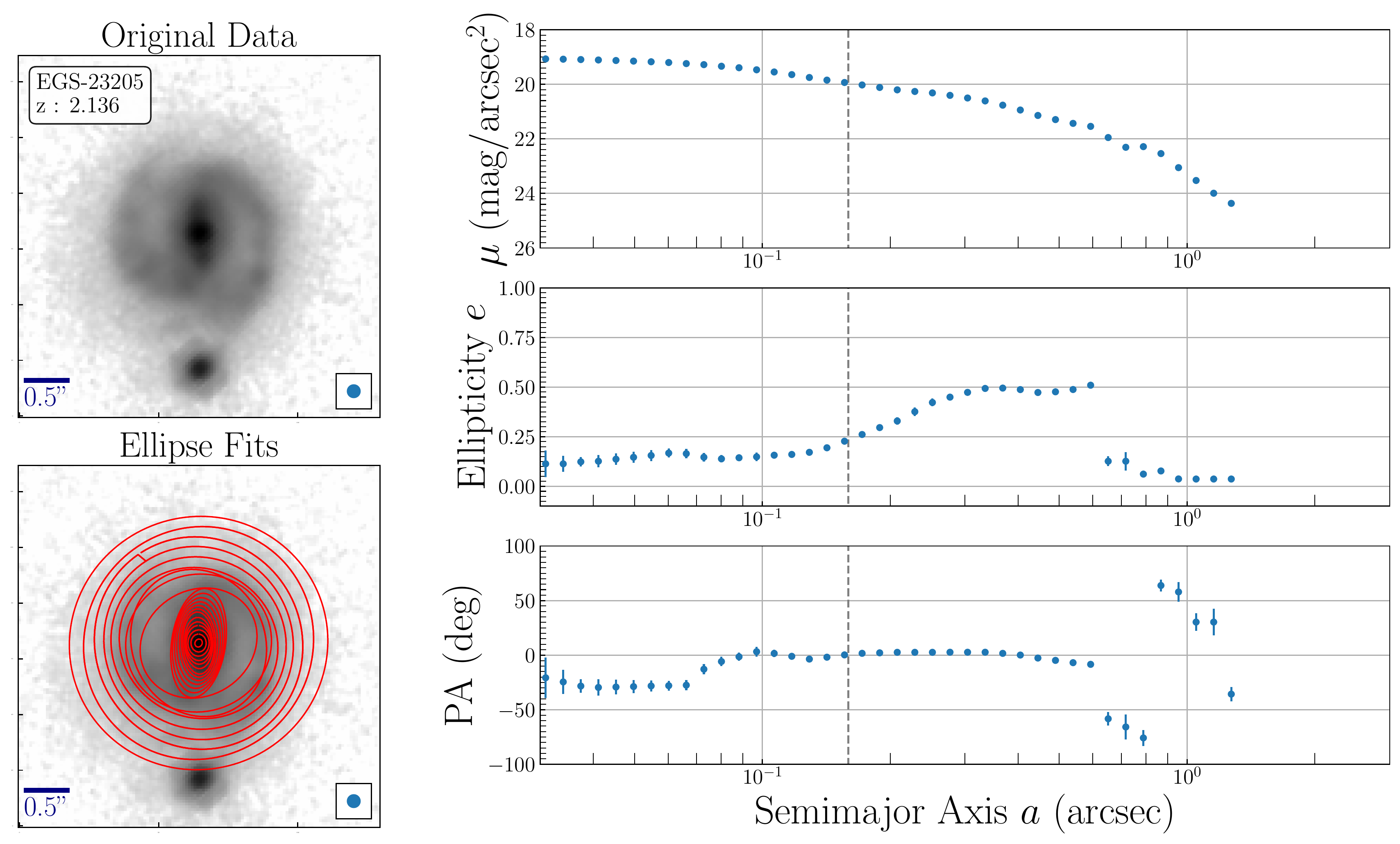}
\includegraphics[width=0.7\textwidth]{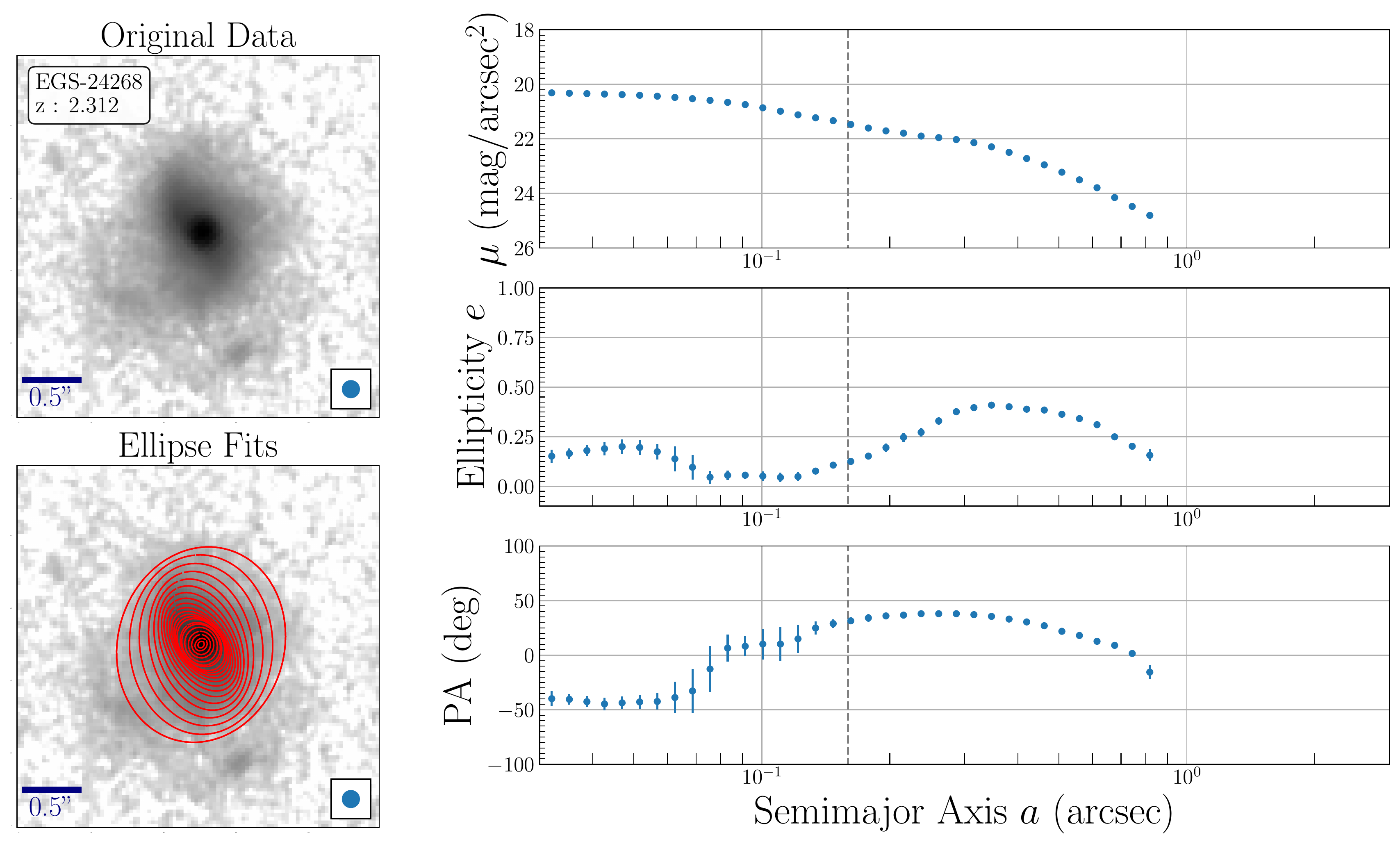} 
\vspace{-1mm}
\caption{
Same as Figure\ref{fig:ellip1} for three other 
example barred galaxies (EGS-26831, EGS-23205, and EGS-24268). See \S~\ref{sec:method} for details.
}
\label{fig:ellip2}
\end{figure*}

Before ellipse-fitting the F444W  images of the 82 putative barred
galaxies,  nearby sources were masked  and the pixel values were
replaced with interpolated values from the nearby region. Then, the ellipse-fitting was done in two steps: 
\begin{enumerate}
\item
We ran ``isophote.Ellipse.fit\_image'' in 
\textsc{Photutils} from Python's astropy package \citep{photutils}
without fixing the center. 
Doing this step, we let the code fully explore the image and return the center of the ellipse for every ellipse fitted. We then determined the center of isophotes in step 2 by measuring the average center of the ellipses fitted to the inner region. 

\vspace{-1mm}
\item
We fixed the center of isophotes at the center measured in step 1
and ran  the same routine ``isophote.Ellipse.fit\_image''. During the fitting, the semi-major
axis grows geometrically by a factor of 1.1 for each step, and the
fitting stops when 
the relative error in the local radial intensity gradient exceeds 0.5 \citep{Busko1996} for two consecutive ellipses or the outermost ellipse extends to the region with low signal-to-noise ratio (SNR $\le$~3). 
From the ellipse fits we generate radial profiles of
surface brightness (SB), ellipticity ($e$), and position angle (PA)
plotted versus the ellipse semi-major axis $a$  (e.g., see Figures
\ref{fig:ellip1} and \ref{fig:ellip2}).
\end{enumerate}

In alignment with best practices in the study of bars, 
we exclude from further consideration of all galaxies 
with large  inclinations ($i >$~60\deg~as inferred from projected axis ratios of the outer ellipse)  as the bar and the outer disk are very hard to 
separate in such systems.
Among the remaining bar candidates, we consider a galaxy to be barred
only if it satisfies the two criteria below (e.g., \citealt{Jogee-etal-2004,Marinova-Jogee2007}):  

\begin{enumerate}

\item 
In the bar-dominated region,  we require the ellipticity $e$ to rise smoothly to a  maximum value $e_{\rm
{max}} >$~0.25, while the PA stays fairly constant along the bar, with some small
variation $\Delta \theta_{\rm 1} $ allowed. We discuss the value of
$\Delta \theta_{\rm 1} $ later in this section.

\vspace{-1mm}
\item
In the region dominated by the outer disk, we require the ellipticity
to drop by at least 0.1  from the bar's maximum ellipticity and
the PA to change by at least 10\deg~from the associated bar PA.
In galaxies where a transition region exists between the end of the
bar and the region dominated by the outer
disk, we apply the above criterion to the outer disk region beyond it.
\end{enumerate}

\begin{figure*}[!t]
\centering
\includegraphics[width= 0.7 \textwidth]{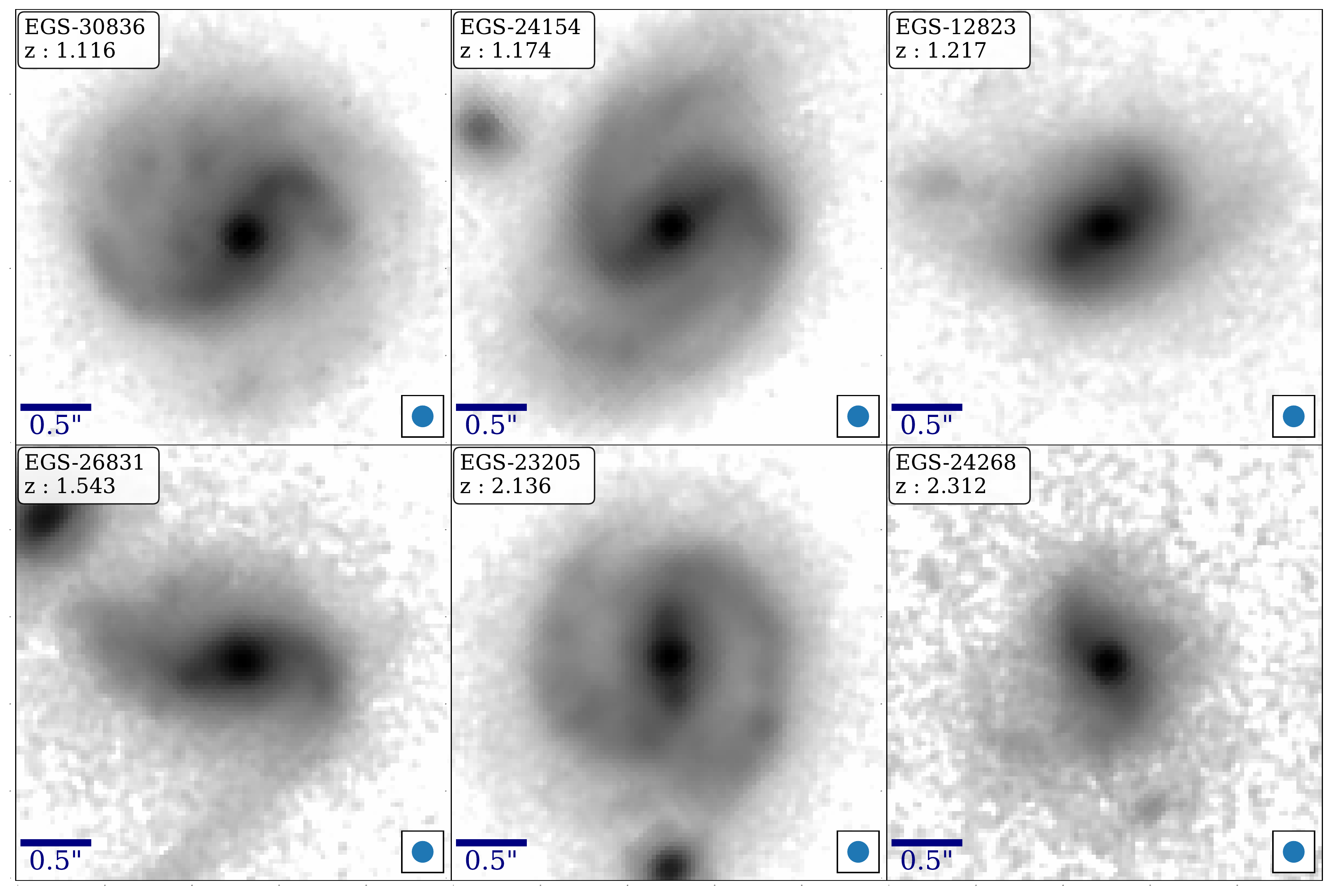}
\vspace{0mm}
\caption{
Montage of {\it{JWST}}  F444W images showing the
rest-frame NIR morphology of 
the  six example barred galaxies presented in this paper.
The bars were identified by applying quantitative criteria to ellipse
fits as outlined in \S~\ref{sec:method}.
The labels in the top left of each figure show the CANDELS ID and redshift of each galaxy.
The galaxies have spectroscopic redshifts of 1.116 , 1.174, 1.217, 1.543,  2.136, and 2.312  
and the last two cases represent the highest redshift bars
quantitatively identified and characterized to date.
The blue circle at the bottom right of each image represents the point
spread function (PSF) FWHM (0\farcs16 corresponding to $\sim$1.3 kpc  at $z\sim$~2), 
and the horizontal bar shows  a 0\farcs5 scale for reference. 
All images are 3\farcs0 $\times$ 3\farcs0 in size.}
\label{fig:barpanel}
\end{figure*}


We refer the reader to \cite{Jogee-etal-2004} and
\cite{Marinova-Jogee2007}  for a
description of the physically motivated justification for 
the above two criteria for bar identification. 
We comment here further on the criteria related to the PA. 
In the first criterion,  we required the PA to stay fairly
constant along the bar, with some small variation $\Delta \theta_{\rm 1} $ allowed. 
The rationale for a relatively constant PA is
that the main {\bf{x}}$_1$  family of  bar-supporting orbits
can be modeled by concentric ellipses with a fairly constant PA   
as a function of  semi-major axis in the bar region \citep{Athanassoula1992a}.  
Many studies do not specify the value they adopt for the allowed
variation  $\Delta \theta_{\rm 1} $, while others use a wide range in $\Delta \theta_{\rm 1} $
from 20\deg~to 40\deg~(e.g., \citealt{Jogee-etal-2004,Marinova-Jogee2007,Olguin-Iglesias-Kotilainen-Chavushyan2020}). 
In this pilot paper, we will only show six examples of robustly identified bar candidates
where the variation $\Delta \theta_{\rm 1}$ of the PA along the bar 
is conservatively low at  $\Delta \theta_{\rm 1} \le $~20\deg ~(see
\S~\ref{sec:resul}).
In our future papers that aim for a more complete
census of bars at $z>1$, we will explore the impact of adopting larger  $\Delta
\theta_{\rm 1} $ values and fine tuning other aspects of the
methodology.


Examples of ellipse fits are shown in Figures~\ref{fig:ellip1}  and 
\ref{fig:ellip2} and discussed in the next section.  
The above two criteria are quite effective in 
separating barred galaxies from  inclined disk galaxies
(e.g., see Appendix Figure~\ref{fig:app1}) and unbarred galaxies 
(e.g., see Appendix Figure~\ref{fig:app1}). 
We also note that short bars will not be identified in the {\it{JWST}} F444W images due
to the loss of spatial resolution. 
We do not expect to robustly identify bars whose semi-major axis is 
less than the PSF of F444W images (0\farcs16 corresponding 
to $\sim$1.3 kpc  at $z\sim$~1--3).

\subsection{Characterization of Bar Length and Maximum Ellipticity
}\label{sec:method-chracterization}

The shape,  
length and stellar mass of a stellar bar are 
important properties that determine  the gravitational torque it exerts
and its impact on the secular evolution of a galaxy.
The main goal of this paper is to identify and demonstrate the
existence  of bars at $z>1$, and the ellipse fits presented in the
previous section are adequate for this purpose as they robustly
identify bars.  
However, for characterizing the strength, shape, and size of bars, 
there are more sophisticated methods  than ellipse fits and 
in our future papers we will explore such methods,  including 
generalized ellipses with a shape parameter
(\citealt{Athanassoula-etal-1990, Gadotti2009a})
and multi-component (bulge, bar, outer disk) decomposition of the light distribution
(e.g., \citealt{Laurikainen-Salo-Buta2005,Laurikainen-etal-2007,Gadotti2009b, Weinzirl-etal-2009}).

In this paper, we focus on the ellipticity and length of the bar based 
on ellipse fits.
We consider the maximum projected ellipticity of the bar $e_{\rm bar}$ 
as one measure of bar strength. 
In the radial profile of projected ellipticity from the ellipse fits
(see Figures~\ref{fig:ellip1}  and \ref{fig:ellip2}), the ellipticity
rises smoothly to a maximum value  in the bar-dominated 
region and we take this maximum value as  $e_{\rm bar}$. 

Different definitions of the bar length are used in the bar community,
including the following: 
(i)  the semi-major axis (sma)  $a_{\rm bar}$  where  the bar ellipticity first
reaches a maximum value along the bar; 
(ii) the sma where the bar ellipticity drops steeply  or by at least 15\% from its maximum
value; 
and (iii)  the sma where  the PA changes from the bar  to the outer
disk.
In this paper we measure bar lengths based on the first definition as the
latter is widely used in many studies  (e.g., \citealt{Athanassoula-Misiriotis2002, Jogee-etal-2004, Marinova-Jogee2007, Menendez-Delmestre-etal-2007}) and can be unambiguously applied to many galaxies. 
However, some studies (e.g., \citealt{Athanassoula-Misiriotis2002, Martinez-Valpuesta-Shlosman-Heller2006})
suggest this definition can underestimate the true bar length.


\begin{deluxetable*}{lcccccc}
\tablewidth{800pt}
\tablenum{1}
\tablecaption{Barred Galaxies at $z>$~1 in the
  Rest-Frame NIR from {\it{JWST}} 
\label{tab:barprop}}
\tablehead{
\colhead{ Galaxy Name} & \colhead{ $z_{\rm spec}$} & \colhead{$e_{\rm bar}$} &
\colhead{$a_{\rm bar}$} & \colhead{$a_{\rm bar}$} & \colhead{log($M_{\star}/M_\odot$)}&\colhead{SFR}\\
\nocolhead{} &\nocolhead{}& \nocolhead{}  & \colhead{(")} &
\colhead{(kpc)} & \nocolhead{} & \colhead{$M_{\odot}$ yr$^{-1}$} 
}
\decimalcolnumbers
\startdata
EGS-30836 & 1.116 (DEEP2 DR4) &  $\sim$ 0.53 & $\sim$~0.51 & $\sim$~4.28 & 10.80 &48.430\\
EGS-24154 & 1.174 (DEEP2 DR4) & $\sim$ 0.52 & $\sim$~0.42 & $\sim$~3.57 & 11.05&45.395 \\ 
EGS-12823 & 1.217 (3D-HST) &  $\sim$ 0.48 & $\sim$~0.38 & $\sim$~3.26 & 10.63 &21.230\\ 
EGS-26831 & 1.543 (MOSDEF) &  $\sim$ 0.49 & $\sim$~0.42 & $\sim$~3.65 & 10.40&74.290 \\
EGS-23205 & 2.136 (3D-HST) &  $\sim$ 0.50 & $\sim$~0.35 & $\sim$~2.95 & 11.29&295.023 \\
EGS-24268 & 2.312 (MOSDEF) &  $\sim$ 0.41 & $\sim$~0.35 & $\sim$~2.91 & 10.16 &112.808\\
\\
\enddata
\tablecomments{Columns are: (1) Galaxy ID from
  \cite{Stefanon-etal-2017}; (2) Spectroscopic redshift and the survey on
  which it is based. The 3D-HST
  redshifts for EGS-12823 and EGS-23205 are based on
  grism spectra in the 3D-HST Survey (\citealt{Brammer-etal-2012, Momcheva-etal-2016}) and are well constrained. 
  The DEEP2 DR4 and MOSDEF spectroscopic redshifts are from 
  \cite{Newman-etal-2013} and \cite{Kriek-etal-2015}, respectively; (3) The
  maximum projected ellipticity  $e_{\rm bar}$ of the stellar bar in the rest-frame NIR based on {\it{JWST}}
  NIRCam F444W images; (4) As in (3), but for the bar projected  semi-major axis
  $a_{\rm bar}$ in arcsec. (5) As in (3),  but for the bar projected semi-major
  axis $a_{\rm bar}$ in kpc. (6) Stellar mass measurements  of the
  host galaxy from  \cite{Stefanon-etal-2017}; (7) Star-formation rate (SFR)
  measurements  of the host galaxy are the best estimate of the total SFR from the value-added catalogs associated with the CANDELS EGS catalogs \citep{Barro-etal-2019}. }
\end{deluxetable*}

\section{Results}\label{sec:resul}

For this pilot paper, we choose to show six examples of  robustly identified 
barred galaxies  that fulfill the following criteria: 
(i)~They  have good ellipse fits of the F444W images (Figures~\ref{fig:ellip1}  and
\ref{fig:ellip2}), unambiguously meet the two bar criteria in
\S~\ref{sec:method-identification}, and 
show a conservatively small variation $\Delta \theta_{\rm 1} \le
$~20\deg~in the PA of ellipses fitted along the bar. As we mentioned in \S~\ref{sec:method-identification}, the variation  $\Delta \theta_{\rm 1}  $  allowed for the PA along
the bar could in general be higher, but we show the most
conservative robust cases of bar here;
(ii)~The bars have moderate to high maximum projected ellipticities ($e_{\rm bar} \sim$~0.41--0.53)  and
they are well resolved with projected semi-major axes $a_{\rm bar} \sim$~0\farcs35--0\farcs51
or  $\sim$~2.9--4.3 kpc  
(Table~\ref{tab:barprop}); 
(iii)~The barred galaxies have a range of published spectroscopic redshifts 
(N. Hathi 2022, private communication), 
at  $z\sim$~1.116, 1.174, 1.217, 1.543,
2.136, and  2.312 (Table~\ref{tab:barprop}),
and include  the two highest redshift
barred galaxies at $z\sim$ 2.136 and  2.312, 
quantitatively identified and characterized to date.

The properties of the six bars and their host galaxies are shown in 
Figure~\ref{fig:barpanel} and Table~\ref{tab:barprop}, and the ellipse fits to their F444W images are shown in Figures~\ref{fig:ellip1}  and \ref{fig:ellip2}. 
In the cases of  EGS-30836,  EGS-24154, EGS-12823 in 
Figure~\ref{fig:ellip1}, as well as EGS-26831, EGS-23205 and EGS-24268 in
Figure~\ref{fig:ellip2}, both bar criteria are  well met: 
the ellipticity rises smoothly  to a maximum in the bar-dominated region while the PA
stays constant within less than 20\deg, and there is a significant drop in ellipticity and
change in PA in the region dominated by the outer disk.

To characterize bar properties, we estimated the maximum  projected
ellipticity ($e_{\rm bar}$) and projected  bar length 
 ($a_{\rm bar}$) of the bar for each galaxy with the method described
 in \S~\ref{sec:method-chracterization}. For our barred galaxies at
 redshifts of $z \sim$ 1.1--2.3, 
the stellar bar has moderate to high maximum projected ellipticities
 ($e_{\rm bar}$)  in the rest-frame NIR  ranging 
from $\sim$~0.41 to 0.53  (Table~\ref{tab:barprop}).
These values overlap with the range of bar projected ellipticities (0.25 to 0.8)
seen in NIR images of $z\sim$~0  bright spirals 
where most ($> 70\%$) bars have $e_{\rm bar}\ge$ 0.4 
(e.g., \citealt{Marinova-Jogee2007, Menendez-Delmestre-etal-2007}). 
Once we have a larger and more complete sample of bars at $z>1$, we can evaluate
whether the distribution of bar strength and ellipticity evolves down
to the present day.  

The projected bar length $a_{\rm bar}$ in the rest-frame NIR ranges from
 $\sim$~2.9--4.3 kpc with angular sizes of  $\sim$~0\farcs35 -- 0\farcs51
(Table~\ref{tab:barprop}).  For the barred galaxies presented in our study, a typical measurement  
error due to ellipse fitting of 0.24--0.42 kpc (0\farcs03--0\farcs05
in angular sizes) on $a_{\rm bar}$ is expected as one cannot 
measure $a_{\rm bar}$ values better than the step size used in
ellipse fitting. 
The range of  $a_{\rm bar}$ values ( $\sim$~2.9--4.3 kpc) in these high
redshift bars  overlaps with the range of stellar bar lengths (1 to 14 kpc)  
seen in NIR images of $z\sim$~0  bright spirals 
where most ($> 75\%$) bars have  have  $a_{\rm bar} \le$~5 kpc (e.g.,
\citealt{Marinova-Jogee2007}). 
However, as mentioned  in \S~\ref{sec:method-chracterization}, we do
not expect to robustly identify bars less than the PSF of F444W images
(0\farcs16 corresponding to $\sim$1.3 kpc  at $z\sim$~1--3), so short
bars are going to be missed in our study.
Additionally, the normalized bar length (ratio of bar length to disk length) is a more meaningful
comparison than using the bar length alone, and we will compute the normalized quantities in future papers.

Our six barred galaxies at $z\sim$~1.1--2.3  have 
published star formation rates (SFRs) $\sim$ 21--295
$M_{\odot}$ yr$^{-1}$  (Table~\ref{tab:barprop}; \citealt{Barro-etal-2019}).
The corresponding specific SFR is $\sim 4 \times 10^{-10}$  to $8 \times
10^{-9}$  yr$^{-1}$, indicating these systems are actively
star-forming. We stress that this result only applies  to the subset
of bars we present here and a wider range of specific SFRs may be
present in the full bar population. 

Our  pilot study of barred galaxies at $z>1$ using high-resolution 
rest-frame NIR images  from  {\it{JWST}} complements 
the many past studies that have used {\it{HST}} data
to explore bars in the rest-frame optical out to $z\sim$~1.2  (e.g.,
\citealt{Elmegreen-Elmegreen-Hirst2004, Jogee-etal-2004, Sheth-etal-2008, Cameron-etal-2010, Melvin-etal-2014}) 
and out to $ z \le$~2.0 \citep{Simmons-etal-2014}.

Our study also complements several other recent {\it{JWST}} studies
that have been submitted or recently accepted for publication. \cite{Jacobs-etal-2022} explore the rest-frame optical morphologies of galaxies at 0.8~$< z <$~5.4 through visual classification of
{\it{JWST}} data, mention  that  ``several sources additionally show
distinct bars'', but provide no further information on the barred galaxies.
\cite{Chen-etal-2022} performs two-dimensional surface brightness
profile fittings of {\it{JWST}} images to explore bulges in
$z \sim$~2 submillimeter galaxies (\citealt{Zavala-etal-2017, Zavala-etal-2018}) and mentions an additional bar component is also needed to improve the fit for EGS-23205 (source 850.025).
Finally, \cite{Ferreira-etal-2022b} explore morphologies of galaxies at 1.5~$< z <$~8 through visual classification of {\it{JWST}} images and focus on disks, spheroids, and peculiar galaxies. They mention in some cases, such as our galaxy  EGS-23205, ``a disk, spiral arms and a bar pops up in the longer wavelength bands", but they do not present a further analysis of the bar.
Our study complements the above studies by using quantitative criteria based on ellipse fits of rest-frame {\it{JWST}} NIR images  to identify bars and to characterize their properties (lengths, ellipticities) and that of their host galaxies. To the best of our knowledge, the two barred galaxies in our pilot study with  spectroscopic redshifts $z \sim$~2.136 and $z \sim$~2.312 are the highest redshift bars quantitatively identified and characterized to date.

\section{Discussion}\label{sec:discu}

When discussing our results, it is important to bear in mind that
our pilot study does not present a full census of bars  and instead, only highlights six examples of  
bars at $z>1$, which have been quantitatively identified  and include
the two highest-redshift bars at $z\sim$ 2.136 and  2.312 known to
date. 
Nonetheless, our present results already allow some interesting conclusions to
be drawn and open up exciting possibilities for future work.



Our finding of well developed bars at $z\sim$~1.1--2.3
with projected semi-major axes of  $\sim$~2.9--4.3 kpc  and 
and projected maximum ellipticities of $\sim$~0.41--0.53 
in the rest-frame NIR (\S~\ref{sec:resul}) 
demonstrates the early onset of such features and 
supports simulations  
where  bars form  early in massive dynamically cold 
disks (e.g., \citealt{Bournaud-Combes2002, Romano-Diaz-etal-2008,
Kraljic-Bournaud-Martig2012, Bonoli-etal-2016, Spinoso-etal-2017,Rosas-Guevara-etal-2022,
Bi-Shlosman-Romano-Diaz2022}).
In a future paper, we will present a census of observable bars at $z
>$~1  and estimate the bar fraction in the rest-frame
NIR out to $z\sim$~3 using F444W images. These images will 
detect obscured bars, but the F444W PSF
(0\farcs16 or  $\sim$1.3 kpc  at  $z\sim$~1--3) will only allow 
the robust identification of bars with length above 1.3 kpc  at
$z\sim$~1--3. 
We will also estimate the rest-frame optical bar fraction out
$z\sim$~4  using F200W and other images. While the rest-frame optical images 
may miss bars impacted by dust and SF, their smaller PSF 
(0\farcs08  or $\sim$~650 pc at $z\sim$~1--3) allows them to detect shorter sub-kpc
bars, which may constitute a significant fraction of the bars in disk galaxies at early
epochs  (e.g., \citealt{Rosas-Guevara-etal-2020, Rosas-Guevara-etal-2022}).
Recent studies in the rest-frame optical identify a large fraction of
disk galaxies at $z\sim 3$ in {\it{JWST}} data  (e.g.,
\citealt{Ferreira-etal-2022a}, \citealt{Kartaltepe-etal-2022}).

The topic of the formation, lifetime, and evolution of bars  is an
area of active research and it depends on the interplay between the
stellar disk, dark matter halo, and gaseous components. 
The properties of the dark matter halo and its exchange of 
angular momentum with stellar or gaseous components impact 
the bar (e.g., \citealt{Athanassoula2003}; \citealt*{Athanassoula-Machado-Rodionov2013}; \citealt{Saha-Naab2013,Sellwood2016,Collier-Shlosman-Heller2018, Beane-etal-2022}).   
The role of gas is complex. While the presence of a massive and dynamically cold disk of stars and gas
favors the onset of $m=2$  bar instabilities (e.g., \citealt{Romano-Diaz-etal-2008, Bournaud-Combes2002,
  Kraljic-Bournaud-Martig2012,Bonoli-etal-2016, Spinoso-etal-2017,
  Rosas-Guevara-etal-2022,Bi-Shlosman-Romano-Diaz2022}),  gas clumps that sink by dynamical friction can heat the stellar
disk  (e.g., \citealt{Shlosman-Noguchi1993}) and
large central gaseous mass concentrations can weaken or destroy
some bars (e.g., \citealt{Bournaud-Combes2002, Shen-Sellwood2004, Athanassoula-Lambert-Dehnen2005, Bournaud-Combes-Semelin2005, Debattista-etal-2006}).

Numerous simulations have also shown that bars can form spontaneously
in isolated disks or be tidally induced (e.g.,
\citealt{Hernquist-Mihos1995,Izquierdo-Villalba-etal-2022}). In that
context, it is interesting to note that many of the  six example
barred galaxies appear to have nearby sources that could be potential companions.
EGS-26831 has a spectroscopic redshift of $z\sim$~1.543 and 
has two potential companions detected in
\cite{Stefanon-etal-2017}: the source to its northeast (partially
visible in Figure \ref{fig:barpanel}) has a similar 
but poorly constrained photometric redshift  (\citealt{Stefanon-etal-2017}) within $\Delta z/(1+z)$
$\sim$~0.065 and is at an angular distance of $\sim$~$1\farcs62$
(corresponding to $\sim$ 14 kpc assuming for $z\sim$~1.543), while the  source 
to the southeast (not visible in  Figure
\ref{fig:barpanel}) has a 
poorly constrained photometric redshift  (\citealt{Stefanon-etal-2017})
within $\Delta z/(1+z)$ $\sim$~0.0056 and is at an angular distance
$\sim$ $2\farcs44$ (corresponding to $\sim$ 21 kpc for $z\sim$~1.543).
Even though those two sources have poorly constrained photometric
redshifts, the interacting features shown in the F444W image
suggest that the sources could potentially be companions of EGS-26831.
For EGS-24154  whose spectroscopic redshift is $z\sim$~1.174, 
the source  to its northeast (partially visible in Figure
\ref{fig:barpanel}) is detected in \cite{Stefanon-etal-2017} 
and could be a potential companion  as it has
a spectroscopic redshift (3D-HST; \citealt{Brammer-etal-2012,Momcheva-etal-2016}) 
within $\Delta z/(1+z)$ $\sim$~0.001 at an angular distance $\sim$
$1\farcs4$  (corresponding to $\sim$ 12 kpc for $z\sim$~1.174). 
For EGS-23205 at  $z\sim$~2.136, the bright source to its south (shown
in Figure \ref{fig:barpanel}) is identified as an X-ray luminous AGN 
with an estimated photometric redshift of $z\sim$~4.1
(\citealt{Kocevski-etal-2022}). Given the difficulty of deriving accurate 
photometric redshifts for luminous AGN, it is unclear how reliable
this redshift is and whether the AGN is a chance projection or a true
companion.
It is also noteworthy that there are faint sources near EGS-30836, EGS-12823
and EGS-24268. Those faint sources are not detected in the CANDELS
EGS catalog \citep{Stefanon-etal-2017}, so we do not have redshifts to
determine whether they are companions, accreted sources or 
overdensities in the disk.
In future papers we will explore the frequency of  tidal interactions in a
larger sample of barred galaxies 
and a control sample of unbarred systems. 

The growth and rich morphological transformation of galaxies
from $z\sim$~4 to today is likely driven by several mechanisms,
including bar-driven secular processes (e.g., \citealt{Sakamoto-etal-1999, Kormendy-Kennicutt2004, Jogee-Scoville-Kenney2005}),  galaxy mergers and tidal
interactions (e.g., 
\citealt{Conselice-etal-2003, Kartaltepe-etal-2007,Jogee-etal-2009, Lotz-etal-2010}), and gas accretion (e.g., \citealt{Katz-etal-2003, Keres-etal-2005, Dekel-Birnboim2006, Faucher-Giguere-Keres2011,Keres-etal-2012}).
Our finding of bars at
$z\sim$~1.1--2.3 (lookback times of 8--10   Gyr) 
also suggests that if these bars survive out to present epochs, 
bar-driven secular processes may operate over a long time  
and have a significant impact on some galaxies  by $z \sim$~0. 
In this context, we note that  \cite{Gadotti-etal-2015} suggest that the bar
in the nearby galaxy NGC 4371 has a formation epoch of $z\sim 2$.

Our subset of barred galaxies at at $z\sim$~1.1--2.3
have SFRs $\sim$ 21--295 $M_{\odot}$ yr$^{-1}$  and specific SFRs $\sim 4
 \times 10^{-10}$  to $8 \times 10^{-9}$  yr$^{-1}$
and are thus actively forming stars. 
Bars drive large gas inflows into the circumnuclear regions via gravitational torques and
shocks and can  lead a phase of high circumnuclear SFR
and potentially depressed or quenched SF in the disk 
(e.g., \citealt{Hunt-Malkan1999, Jogee-Scoville-Kenney2005, Masters-etal-2010, Masters-etal-2012,Khoperskov-etal-2018,  George-Subramanian2021}). We will
explore more fully the impact of bars on galaxy SFRs in a future paper
where we will incorporate the upcoming additional CEERS pointings and
make a statistical comparison of the SFRs of barred and unbarred
systems.

\section{Summary}

The exploration of stellar bars out to early cosmic times is essential
for understanding the evolution of galaxies as bars play a critical role
in driving the secular evolution of  galaxies.
Stellar bars can be effectively mapped in rest-frame NIR images, which
trace the underlying stellar mass and are less impacted by dust and
star formation  than rest-frame UV or optical images.
In this pilot study we conduct the first quantitative identification
and  characterization of  stellar bars at $z>1$  in high-resolution
rest-frame NIR images by analyzing {\it{JWST}} F444W images 
in the first epoch of imaging from the CEERS survey.
We focus on  a sample of 348 galaxies at redshifts 1~$\le z \le $~3, with 
stellar mass $M_{\ast} \ge 10^{10} M_{\odot}$ and CANDELS multi-wavelength data.

The {\it{JWST}} F444W images allow us achieve  a high-resolution  (0\farcs16
corresponding to $\sim$~1.3 kpc at $z\sim$~1--3) at rest-frame NIR wavelengths
to quantitatively identify and characterize bars at  $z>1$.
We identify stellar bars by performing a first-pass visual
classification, followed by ellipse fits and the application of
physically motivated quantitative criteria to the ellipse fits. 
For this pilot study we present 
six examples of robustly identified bars at  $z>1$ with
spectroscopic redshifts, including the two highest redshift bars  at  $z\sim$~2.136  and
2.312, quantitatively identified and characterized to date. 
Our study complements {\it{HST}}  studies in the last two decades 
that have mainly traced bars in the rest-frame optical out to $z\sim$~1.

The examples of stellar bars at $z\sim$~1.1--2.3 presented  in our study
have projected semi-major axes of  $\sim$~2.9--4.3 kpc  and 
moderate to high projected maximum ellipticities
of $\sim$~0.41--0.53 in the rest-frame NIR,  indicating
they are already fairly strong and well developed at these early cosmic times.
The barred host galaxies have stellar masses  $\sim 1 \times
10^{10}$  to $2 \times 10^{11}$ $M_{\odot}$, 
star formation rates of  $\sim$ 21--295 $M_{\odot}$ yr$^{-1}$, and
several have potential nearby companions.
Our finding of bars at $z\sim$~1.1--2.3  demonstrates the early onset
of such instabilities and supports simulations  
where  bars form  early in massive dynamically cold  disks. 
It also suggests that if  these bars at lookback times of 8--10
Gyr  survive out to present epochs, bar-driven secular processes may operate  over a long time 
and have a significant impact on some galaxies  by $z \sim$~0.

This pilot study only presents six examples of robustly identified bars at
$z>1$ in the rest-frame NIR. We do not  present here a full census of
all observable bars at $z >$~1 and the associated statistical distribution of their properties. 
In future papers that will incorporate the upcoming additional
six CEERS pointings, we will present such a census,  estimate the
rest-frame optical and  NIR bar fraction,  
 and explore the relationship between  
bars and galaxy properties (SF, bulges, AGN, and presence of  
companions) using a control sample of unbarred galaxies.  
 

 
Version v0.07 of the CEERS data reduction was used in this work. The
full set of the latest CEERS data products can be found at MAST via 
\dataset[https://doi.org/10.17909/z7p0-8481]{https://doi.org/10.17909/z7p0-8481}.

\begin{acknowledgements}
We thank the entire {\it JWST} team, including the engineers for 
making possible this wonderful over-performing telescope, the 
commissioning team for obtaining these early data, and the pipeline 
teams for their work over the years building and supporting the 
pipeline.  
YG and SJ acknowledge support from the Roland K. Blumberg Endowment in  
Astronomy and Heising-Simons Foundation grant 2017-0464. MBB and SLF acknowledge support from NASA through STScI ERS award  JWST-ERS-1345.  
The authors acknowledge the Texas Advanced Computing Center (TACC) at The University of Texas at Austin for providing HPC and visualization resources that have contributed to the research results reported within this paper. URL: http://www.tacc.utexas.edu
 
\end{acknowledgements}

\software{Astropy \citep{astropy},
          \textsc{Photutils} \citep{photutils},
          SciPy \citep{scipy},
          \textsc{Source Extractor} \citep{Bertin-Arnouts96},
          STScI \textit{JWST} Calibration Pipeline (\url{jwst-pipeline.readthedocs.io})
}

\pagebreak
\bibliographystyle{aasjournal}
\bibliography{kg-sj}

\appendix

\begin{figure*}[h]
\centering

\includegraphics [width=0.7 \textwidth]{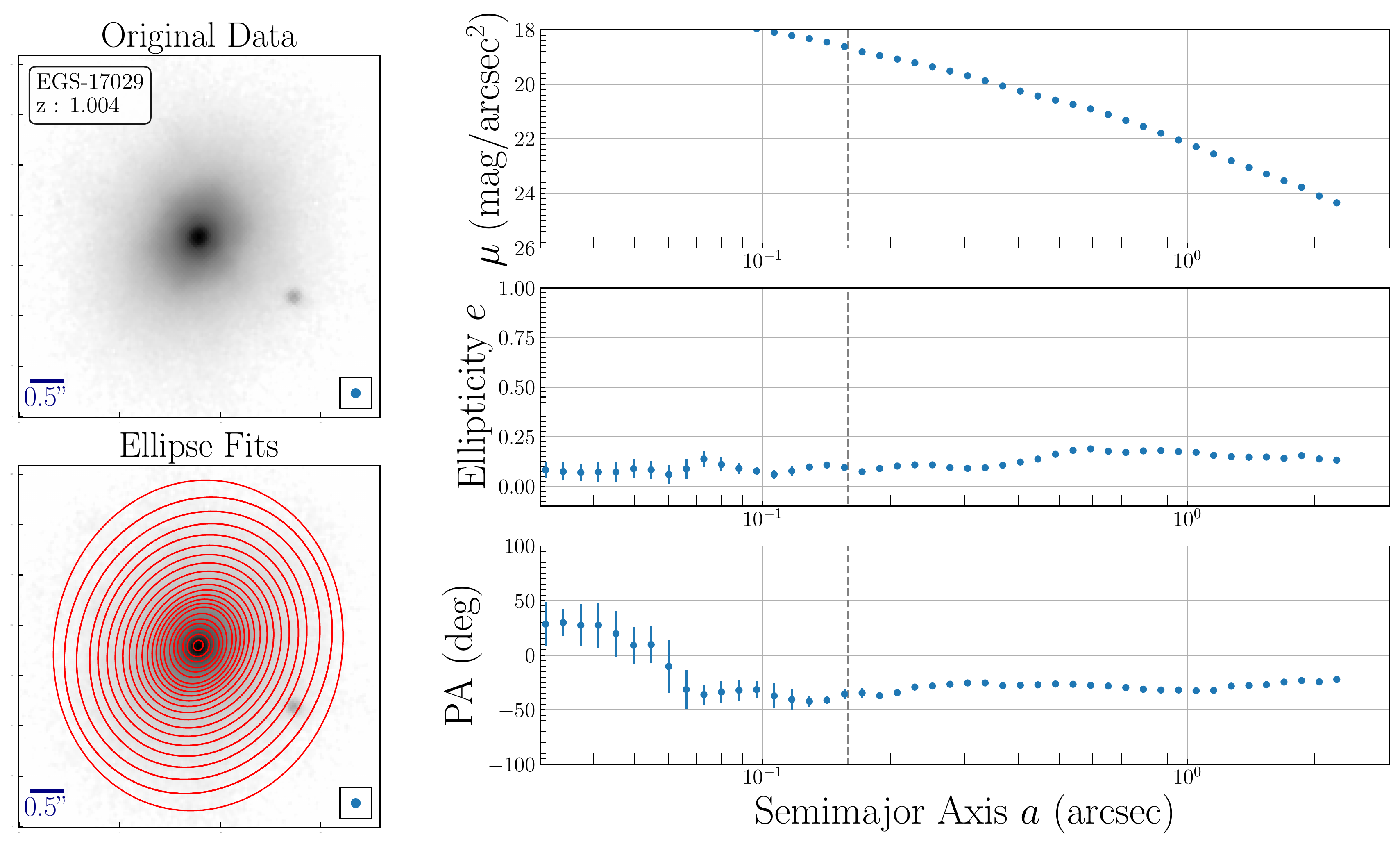}
\includegraphics [width=0.7 \textwidth]{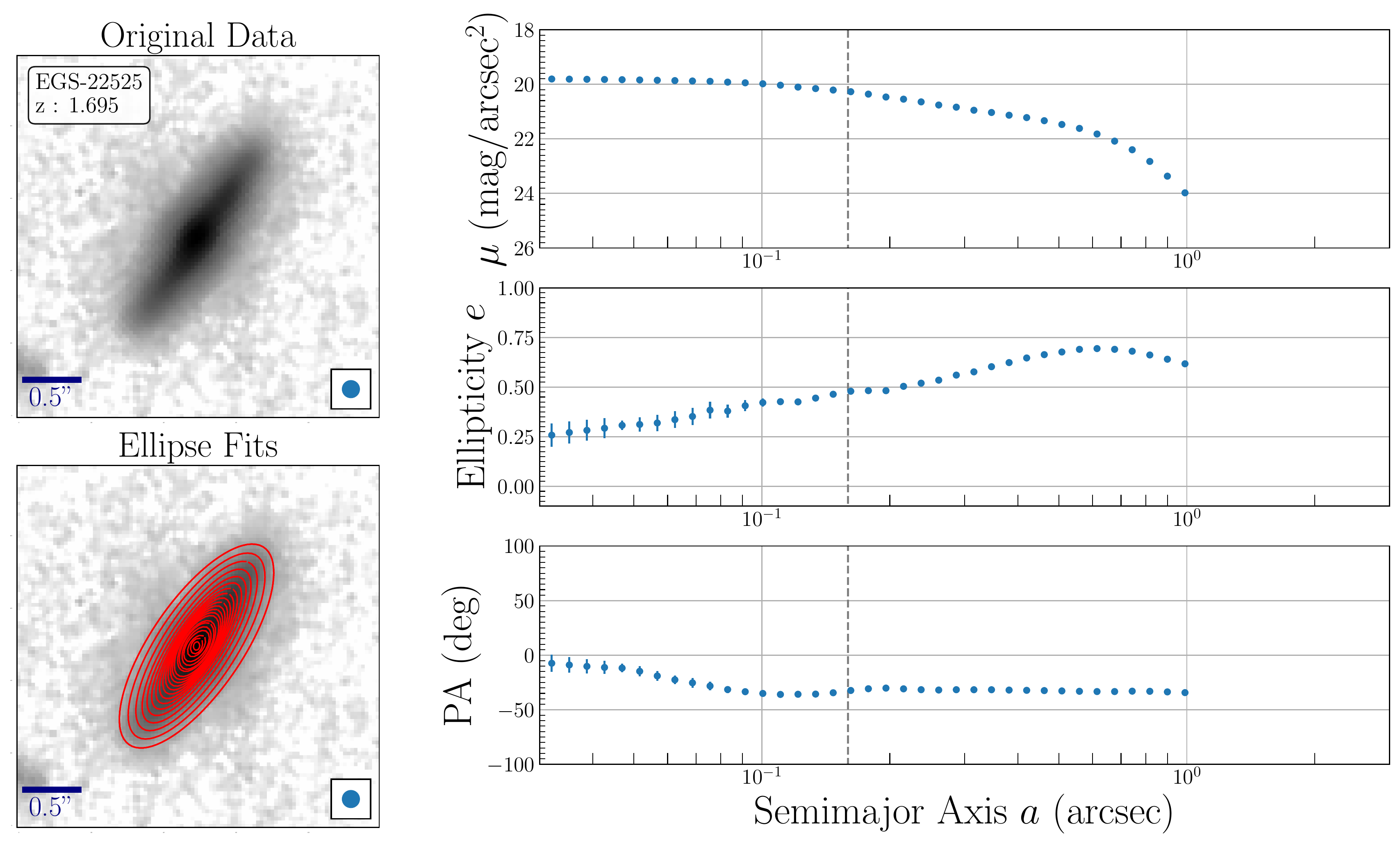}

\vspace{-1mm}
\caption{
Ellipse fits to the {\it{JWST}} NIRCam F444W image of 
an unbarred face-on disk galaxy (EGS-17029) and an inclined disk galaxy (EGS-22525).
The left panel for each galaxy shows 
the F444W image alone (top) and then with the ellipse fits superposed
(bottom). The blue circle at the bottom right of each image represents
the PSF FWHM (0\farcs16 corresponding to $\sim$ 1.3 kpc  at
$z\sim$~1-3), and the horizontal bar shows  a 0\farcs5  scale for reference. Size of each image is adjusted with respect to the size of the source, and ranges from 4\farcs5 $\times$ 4\farcs5 to 3\farcs0 $\times$ 3\farcs0.
The right panel for each galaxy shows
the radial profiles of surface brightness ($\mu$), ellipticity ($e$),
and position angle (PA)  versus semi-major axis $a$ derived from the
ellipse fits. The vertical dashed line
represents the F444W PSF (0\farcs16). The profiles do not show the characteristic bar
signatures that meet our criteria in \S~\ref{sec:method-identification}.
}
\label{fig:app1}
\end{figure*}

\suppressAffiliationsfalse
\allauthors
\end{document}